\documentclass[twocolumn,showpacs,aps]{revtex4}
\usepackage{amssymb}

\usepackage[english]{babel}

\usepackage{amsmath,amssymb,amsxtra,amsfonts}
\usepackage{graphicx}
\usepackage{subfigure}

\usepackage{txfonts}

\begin{document}

\title{Cosmological constraints on holographic dark energy models under the
energy conditions}

\author{Ming-Jian Zhang${}^1$, Cong Ma${}^1$, Zhi-Song Zhang ${}^2$, Zhong-Xu Zhai${}^3$,
, Tong-Jie Zhang${}^{1,4}$ } \email{tjzhang@bnu.edu.cn}

  \affiliation{${}^1$ Department of Astronomy, Beijing Normal University, Beijing, 100875, China}
  \affiliation{${}^2$ Department of Aerospace Engineering, School of Astronautics, Harbin Institute of Technology (HIT), Harbin Heilongjiang, 150001, China }
  \affiliation{${}^3$ Department of Physics, Institute of Theoretical Physics, Beijing Normal University, Beijing, 100875, China}
  \affiliation{${}^4$ Center for High Energy Physics, Peking University, Beijing, 100871, China}

\begin{abstract}
We study the holographic and agegraphic dark energy models without
interaction using the latest observational Hubble parameter data
(OHD), the Union2.1 compilation of type Ia supernovae (SNIa), and
the energy conditions. Scenarios of dark energy are distinguished by
the cut-off of cosmic age, conformal time, and event horizon. The
best-fit value of matter density for the three scenarios almost
steadily located at $\Omega_{m0}=0.26$ by the joint constraint. For
the agegraphic models, they can be recovered to the standard
cosmological model when the constant $c$ which presents the fraction
of dark energy approaches to infinity. Absence of upper limit of $c$
by the joint constraint demonstrates the recovery possibility. Using
the fitted result, we also reconstruct the current equation of state
of dark energy at different scenarios, respectively. Employing the
model criteria $\chi^2_{\textrm{min}}/dof$, we find that conformal
time model is the worst, but they can not be distinguished clearly.
Comparing with the observational constraints, we find that SEC is
fulfilled at redshift $0.2 \lesssim z \lesssim 0.3$ with $1\sigma$
confidence level. We also find that NEC gives a meaningful
constraint for the event horizon cut-off model, especially compared
with OHD only. We note that the energy condition maybe could play an
important role in the interacting models because of different
degeneracy between $\Omega_m$ and constant $c$.

\end{abstract}

 \maketitle

\section{Introduction}
\label{introduction}

Several independent cosmological observations probed a phenomena of
an accelerated expansion of the universe.  Examples include the Type
Ia supernovae (SNIa) observations \citep{riess1998supernova}, large
scale structure \citep{tegmark2004cosmological}, and cosmic
microwave background (CMB) anisotropies  \citep{spergel2003wmap}.
Generally, dark energy models or some modified gravity are approved
to be theoretical explanations of this acceleration. For the dark
energy, however, many variants are listed as one candidate. They are
usually in the form of cosmological constant, quintessence
\citep{peebles1988cosmology}, K-essence
\citep{armendariz2000dynamical}, tachyon
\citep{padmanabhan2002accelerated}, phantom
\citep{caldwell2003phantom}, ghost condensate
\citep{piazza2004dilatonic}, quintom \citep{feng2005wang} and the
holographic dark energy which will be investigated in this paper.
Observationally, many of them can fit well with current
observations.

Historically, the holographic dark energy is addressed to alleviate
or remove the cosmological constant problems
\citep{peebles1988cosmology,ozer1986possible,sahni2000case}. From
the origin, it dates from the holographic principle and inspire from
the Bekenstein entropy bound. In an effective quantum field theory,
the total entropy in a box of size $L$ with UV cut-off $\Lambda$
relating to the quantum zero-point energy
\citep{1995hologram,2001holographic} should satisfy the relation
$
    L^{3} \Lambda^{3} \leqslant S_{BH} \equiv \pi M^{2}_{p} L^{2},
$ where $S_{BH}$ is the entropy of a black hole within the same size
as $L$, and $M_{p} \equiv 1 / \sqrt{8 \pi G}$ is the reduced Planck
mass.  Cohen et al. \cite{cohen1999effective} suggested that a short
distance cut-off in quantum field theory is related to a long
distance cut-off, due to the limit set by formation of a black hole.
If $\rho_{\Lambda}$ is the quantum zero-point energy density caused
by a short distance cut-off, the total energy in a region of size
$L$ should not exceed the mass of a black hole of the same size,
namely, $L^{3}\rho_{\Lambda} \leqslant L M^{2}_{p}$. The largest $L$
allowed is the one saturating this inequality. The energy density
should, therefore, satisfy
\begin{equation}  \label{zeropointenergy}
    \rho_{\Lambda} = 3 c^{2} M^{2}_{p} L^{-2},
\end{equation}
where $c$ is a dimensionless constant introduced for convenience,
but indicating the abundance of matter or dark energy component. It
should not be confused with the speed of light $c(\textrm{light})$.
If considering the cut-off $L$ on the cosmology as the size of the
universe, the holographic dark energy finally was born. Initially,
the cut-off $L$ is construed as the Hubble horzion $H^{-1}$
\citep{hovrava2000probable,thomas2002holography,hsu2004entropy}.
However, it can not deduce the accelerating expansion of the
universe \citep{hsu2004entropy}. Subsequently, some length scales
are proposed, such as the particle horizon, future event horizon.
This kind of models are usually named the holographic dark energy.
Conversely, some agegraphic dark energy models corresponding to
temporal scales also come into being in following years, e.\ g.\ the
age of the universe or the conformal time. Nevertheless, the latter
is also accused of classically unstable and worse fitting result
\citep{kim2008instability}. This is mainly because the constant $c$
is difficult to determinate in such models. On the other hand,
instead of the $\rho_{\Lambda}$ as the \textit{total} energy, many
modified dark energy models are proposed \citep{wei2009modified}.

The confrontation of theories and data, however, is not all one may
perform in the face of the proliferation of dark energy proposals.
One of the many interesting approaches makes use of the so-called
{\em energy conditions}.  These conditions were introduced by
Hawking et al. \cite{hawking1975large} as coordinate-invariant
inequality constraints on the energy-momentum tensor that appears in
the Einstein field equation. Due to their simplicity and model
independence, the energy conditions are frequently discussed in the
general context of gravitation \citep{wald1984general}.  Of the many
proposed energy conditions, the ones we employ in this paper are the
null, weak, strong, and dominant energy conditions (abbreviated
respectively as NEC, WEC, SEC, and DEC). The energy conditions in
the setting of a homogeneous and isotropic
Friedmann-Robertson-Walker (FRW) universe summarized by Carroll
\cite[][Chapter 4]{carroll2004spacetime} can be expressed in terms
of the energy density $\rho$ and the pressure $p$ as following:
\begin{equation} \label{ec}
    \begin{array}{lll}
    \mbox{NEC}: \  &\, \rho + p \geqslant 0 \;,  &   \\
    \\
    \mbox{WEC}: \ \  & \rho \geqslant 0 &
    \ \mbox{and} \quad\, \rho + p \geqslant 0 \;,  \\
    \\
    \mbox{SEC}:   & \rho + 3p \geqslant 0 &
    \ \mbox{and} \quad\, \rho + p \geqslant 0 \;, \\
    \\
    \mbox{DEC}:    & \rho \geqslant 0  & \ \mbox{and} \; -\rho \leqslant p
    \leqslant \rho \;.
    \end{array}
\end{equation}

The energy conditions has been useful in discussing the general
property of fluid models \citep[][and references
therein]{wu2012reconstructing}.  The energy conditions were applied
to cosmology by Visser
\cite{visser1997energy,visser1997general,visser2000energy}.  It has
been shown that constraints on a variety of cosmological variables
or parameters can be predicted from the energy conditions, such as
the Hubble parameter, luminosity and angular diameter distances,
lookback time \citep{visser1997general}, total density parameter
$\Omega(z)$, energy density $\rho(z)$, and pressure $p(z)$
\citep{cattoen2008cosmodynamics}.  In a word, the energy conditions
have been an useful tool for our understanding of the Universe's
evolution.

Unlike first-principle laws of physics, the energy conditions are
not expected to hold {\it a priori}, nor are they found to do so
from data. It has been shown that the WEC and DEC are fulfilled for
$z\leq 1$ and $z\lesssim 0.8$ by supernova data with $3\sigma$
confidence levels, respectively \citep{lima2008energy}.
Nevertheless, SEC violation is a typical trait of a positive
cosmological constant $\Lambda$ \citep{li2011dark} and other dark
energy models \citep{schuecker2003observational}.  It is also said
that the WEC violate the quantum field theory due to arbitrarily
negative renormalized energy density may occur at some points of
spacetime \citep{epstein1965}.  If extended regions of large
negative energy density emerge in the nature, exotic and possibly
undesirable phenomena may be allowed, ranging from violations of the
second law of thermodynamics and cosmic censorship to the creation
of time machines and warp drives \citep{fewster1998bounds}.

Recently, Wu et al. \cite{wu2012reconstructing} studied the
likelihood of energy condition violations in the history of the
universe. They found that the data suggest a fulfillment of null and
dominant energy conditions, but a violation of strong energy
condition, especially for low redshift ($z \leqslant 0.3$). They
also noted the difficulty of assessing the possibility of SEC
violation at the high redshift. Moreover, their result for SEC hints
at a recent transition from deceleration to acceleration  with the
transition redshift $z \approx 0.5$ under the ignorance of bias at
the high redshift. Therefore, the SEC fulfillment appears disfavored
as a test for cosmological models.  On the contrary, a dark energy
model is expected to reproduce its violation for recent eras of
cosmic evolution.

This paper is organized as follows: In Sec.\ \ref{model} the basic
expansion rate for general holographic and agegraphic dark energy
models (collectively abbreviated as HDE) is derived and calculated for
different choices of the IR cut-off.  Proceeding to Sec.\ \ref{observation}
we subject the models to our data constraints and energy condition analyses.
Our main results and discussions are presented in Sec.\ \ref{Conclusions}.

\section{Holographic and agegraphic dark energy models}
\label{model}

In this section we consider the HDE models without interaction
\citep{wu2008current} between the dark energy and matter (including
both baryon matter and cold dark matter). The Friedmann equation for
a spatially flat FRW model ignoring the radiation reads
\citep{li2004model,zhang2007constraints}
\begin{equation}  \label{total Friedmann}
    \rho_{m} + \rho_{\Lambda} = 3 M^{2}_{p} H^{2}.
\end{equation}
By introducing $\Omega_{m} = \rho_{m} / (3 M^{2}_{p} H^{2})$ and
$\Omega_{\Lambda} = \rho_{\Lambda} / (3 M^{2}_{p} H^{2})$, the Friedmann
equation can also be cast as  $\Omega_{m} + \Omega_{\Lambda} = 1$.
With the dark energy pressure $p = \omega \rho$ and cold dark
matter pressure $p = 0$, the total pressure $p$ is given by
\begin{equation} \label{total press}
    p = 3 M^{2}_{p} \omega c^{2} L^{-2},
\end{equation}
where $\omega$ is called the equations of state (EoS) parameter. Following
previous works
\citep{li2004model,setare2006interacting,zhai2011constraints,chen2011yun},
the continuity equations for dark energy and matter respectively are
\begin{equation}\label{Eq:exchange}
  \dot{\rho}_{m} + 3 H (1 + \omega_{m}) \rho_{m} = 0, \quad
  \dot{\rho}_{\Lambda} + 3 H (1 + \omega_{\Lambda}) \rho_{\Lambda} = 0,
\end{equation}
where a dot above denotes the derivative with respect to the cosmic time
$t$. Connecting Eqs.\ (\ref{total Friedmann}) and (\ref{Eq:exchange}) we can
obtain
\begin{equation}\label{plus rho}
    2 M^{2}_{p} \dot{H} + 3 M^{2}_{p} H^{2} + \omega_{m} \rho_{m} +
    \omega_{\Lambda} \rho_{\Lambda} = 0.
\end{equation}
By virtue of $\omega_{m} = 0$, the EoS of HDE, with the help of the
second equation of Eq.\ (\ref{Eq:exchange}), can be expressed as
\begin{eqnarray} \label{lambda eos}
    \omega_{\Lambda} &=& -1 - \frac{1}{3 H}
    \frac{\dot{\rho}_{\Lambda}}{\rho_{\Lambda}} = -1 - \frac{1}{3 H
    \Omega_{\Lambda}} \left(\frac{2 \Omega_{\Lambda}}{H} \frac{dH}{dt} +
    \frac{d\Omega_{\Lambda}}{dt} \right) \nonumber\\
    &=& -1 - \frac{2}{3} \left( \frac{d\ln H}{d\ln a} + \frac{1}{2} \frac{d\ln
    \Omega_{\Lambda}}{d \ln a} \right).
\end{eqnarray}
Applying the EoS of HDE, Eq.\ (\ref{plus rho}) reduces to
\begin{equation} \label{lnH}
    \frac{d\ln H}{d\ln a} - \frac{1}{2} \frac{\Omega_{\Lambda}}{(1 -
    \Omega_{\Lambda})} \frac{d\ln \Omega_{\Lambda}}{d\ln a} + \frac{3}{2} = 0.
\end{equation}
Meanwhile, we get the Hubble parameter $H(z) = H_{0} E(z)$ with
Eq.\ (\ref{lambda eos}) and Friedmann equation Eq.\ (\ref{total Friedmann}),
where the expansion rate $E(z)$ is given by \citep{zhai2011constraints}
\begin{equation}\label{Eq:expansionrate}
  E^{2}(z) = \Omega_{m0} (1 + z)^3 + \Omega_{\Lambda0} \exp \left[
  3 {\int_{0}^{z} \frac{1 + \omega_{\Lambda}(z')}{1 + z'} dz'} \right].
\end{equation}
Here the subscript ``0'' denotes the present value of a quantity.

To further quantify a model, a choice of the IR cut-off is needed.
Scenarios summarized by Chen et al. \cite{chen2011yun} contains the
Hubble horizon, the particle horizon, the event horizon, the age of
the universe and the conformal time
\citep{xu2010time,zhai2011constraints}. However, an accelerating
expansion of the universe cannot be achieved when the Hubble horizon
or the particle horizon is chosen as the IR cut-off.  In this paper,
we mainly aim at the last three scenarios.  Recently, Zhai et al.
\cite{zhai2011constraints} investigated four of them and found that
the event horizon is more preferable than Hubble horizon scenario.
However, both two temporal scenarios better recovers the
$\Lambda$CDM model than the spatial scenarios.  In our following
calculations we choose the natural units of the speed of light
$c(\textrm{light})=1$. Our final results will be presented in
dimensionless quantities.

\subsection{Cosmic age cut-off}
\label{sec:cosmicage}

The first scenario, the cosmic age cut-off, is defined as
\begin{equation}\label{age horizon}
  t_{\Lambda} = \int_{0}^{t} dt' = \int^{a}_{0} \frac{da'}{H a'}.
\end{equation}
In this case, the age of universe is considered as a time scale. The
corresponding spatial scale is obtained after multiplication by the
speed of light $c(\textrm{light})$. According to Eq.\
(\ref{zeropointenergy}) the resulting dark energy density is
\begin{equation}
    \rho_{\Lambda} = 3 c^{2} M^{2}_{p} t_{\Lambda}^{-2}.
\end{equation}
With the dark energy density parameter $\Omega_{\Lambda} =
\rho_{\Lambda} / (3 M^{2}_{p} H^{2})$ and the definition Eq.\
(\ref{age horizon}), we find
\begin{equation} \label{age int}
    \int^{a}_{0} \frac{d\ln a'}{H} = \frac{c}{H}
    \sqrt{\frac{1}{\Omega_{\Lambda}}}.
\end{equation}
Taking the derivative with respect to $\ln a$, a differential equation is
derived
\begin{equation}\label{age lnH}
    \frac{d\ln H}{d\ln a} + \frac{1}{2} \frac{d\ln \Omega_{\Lambda}}{d\ln a} +
    \frac{\sqrt{\Omega_{\Lambda}}}{c} = 0.
\end{equation}
Then, from Eqs.\ (\ref{lnH}) and (\ref{age lnH}), the
evolution of $\Omega_{\Lambda}$ can be found to satisfy
\begin{equation}
    \label{Lambda evolution age}
    \frac{d\Omega_{\Lambda}}{dz} = -2 \Omega_{\Lambda} (1 - \Omega_{\Lambda})
    \left( \frac{3}{2} - \frac{\sqrt{\Omega_{\Lambda}}}{c} \right) (1 +
    z)^{-1}.
\end{equation}
Meanwhile, the EoS for HDE can be obtained from Eqs.\ (\ref{lambda eos}) and
(\ref{age lnH})
\begin{equation}\label{age eff}
    \omega_{\Lambda} = \frac{2}{3 c} \sqrt{\Omega_{\Lambda}} - 1.
\end{equation}

Accelerated expansion requires $c > \sqrt{\Omega_{\Lambda}}$ in
order to satisfy $\omega < -1/3$.  When the constant $c$ approaches infinity,
the $\Lambda$CDM model is recovered.  The expansion rate under this scenario
can eventually be determined by Eqs.\ (\ref{Eq:expansionrate}) and (\ref{age
eff}).

\subsection{Conformal time cut-off}

The second scenario we consider is the conformal time as the IR
cut-off. It is in the form
\begin{equation}
    \eta_{\Lambda} = \int_{0}^{t} \frac{dt'}{a} = \int^{a}_{0} \frac{da'}{H
    a'^{2}},
\end{equation}
which is the total comoving distance that light could travel
\citep{myung2009origin}.  In this case, the conformal time is
considered as a temporal scale, and we can again convert it to a spatial scale
to be used as the IR cut-off $L$.  Analogous to Section \ref{sec:cosmicage} we
obtain a differential expression of $\Omega_{\Lambda}$  with respect to $\ln
a$:
\begin{equation}\label{conformal lnH}
    \frac{d\ln H}{d\ln a} + \frac{1}{2} \frac{d\ln \Omega_{\Lambda}}{d\ln a} +
    \frac{\sqrt{\Omega_{\Lambda}}}{a c} = 0.
\end{equation}

We can go further with the aid of Eq.\ (\ref{lnH}):
\begin{equation}
    \label{Lambda evolution conformal}
    \frac{d\Omega_{\Lambda}}{dz} = -2 \Omega_{\Lambda} (1 - \Omega_{\Lambda})
    \left[ \frac{3}{2} (1+z)^{-1} - \frac{\sqrt{\Omega_{\Lambda}}}{c} \right].
\end{equation}
The EoS for HDE can be obtained from
Eqs.\ (\ref{lambda eos}) and (\ref{conformal lnH})
\begin{equation}\label{conformal eff}
    \omega_{\Lambda} = \frac{2}{3} \frac{\sqrt{\Omega_{\Lambda}}}{c} (1 + z) -
    1,
\end{equation}
which corresponds to an acceleration when $c >
\sqrt{\Omega_{\Lambda}} (1+z)$. With $c \rightarrow \infty$ it will
also recover the $\Lambda$CDM model.  The corresponding expansion
rate Eq.\ (\ref{Eq:expansionrate}) of this scenario can also be
solved.

\subsection{Event horizon cut-off}

For our last model, we consider the event horizon \citep{2011Bhattacharya}
given by
\begin{equation} \label{event horizon}
    R_{E} = a \int^{\infty}_{t} \frac{dt'}{a(t')} = a \int^{\infty}_{a}
    \frac{da'}{H a'^{2}},
\end{equation}
which is the boundary of the volume a fixed observer may eventually
observe.  We now identify this scale as the cut-off length $L$ as it appears in
Eq.\ (\ref{zeropointenergy}).  With the introduction of dark energy density
parameter $\Omega_{\Lambda}$, and performing the same analysis, we obtain the
relation from Eq.\ (\ref{event horizon})
\begin{equation}  \label{event int}
    \int^{\infty}_{a} \frac{d\ln a'}{H a'} = \frac{c}{H a}
    \sqrt{\frac{1}{\Omega_{\Lambda}}}.
\end{equation}
Taking the derivative with respect to $\ln a$ of Eq.\ (\ref{event int}), we
get a differential equation
\begin{equation} \label{event lnH}
    \frac{d\ln H}{d\ln a} + \frac{1}{2} \frac{d\ln \Omega_{\Lambda}}{d\ln a} =
    \frac{\sqrt{\Omega_{\Lambda}}}{c} - 1.
\end{equation}
In this manner we further obtain the evolution of $\Omega_{\Lambda}$
from the above Eqs. (\ref{lnH}) and (\ref{event lnH})
\begin{equation}
    \label{Lambda evolution event}
    \frac{d\Omega_{\Lambda}}{dz} = -2 \Omega_{\Lambda} (1 - \Omega_{\Lambda})
    \left( \frac{1}{2} + \frac{\sqrt{\Omega_{\Lambda}}}{c} \right) (1 +
    z)^{-1}.
\end{equation}
Same as the other scenarios but for Eqs.\ (\ref{lambda eos}) and (\ref{event
lnH}), the EoS for HDE is found to be
\begin{equation}\label{event eff}
    \omega_{\Lambda} = -\frac{1}{3} \left( \frac{2}{c} \sqrt{\Omega_{\Lambda}}
    + 1 \right).
\end{equation}

Obviously, the acceleration condition $\omega < -1/3$ is satisfied
as long as $c > 0$. With $c \rightarrow \infty$, we have $\omega
\rightarrow -1/3$.  This is distinguished from the previous two
scenarios.  As with the other models, the Hubble expansion rate
$E(z)$  follow easily.

\section{Observational constraints and the energy conditions}
\label{observation}

Holographic and agegraphic dark energy models have been examined
using various astronomical observations, such as SNIa
\citep{huang2004supernova} and CMB \citep{enqvist2005searching},
among others. Indeed, constraint on HDE from OHD is also performed
\citep{yi2007constraints,zhai2011constraints}. Using the OHD and
$\chi^{2}$ statistics, the authors gave the best-fit  values of
parameters $\Omega_{m0}$ and $c$, suggesting that $c < 1$ is
favored. Various estimations of the $c$ parameter has been made
\citep{huang2004supernova,kao2005cmb,zhang2007constraints}, however
there appears to be little consensus about its precise range.  The
constant $c$ is a key parameter in the HDE model which represents
the proportion of dark energy component as shown in Eq.\
(\ref{zeropointenergy}). In this paper, we therefore try to use the
current more abundant OHD and SNIa compilation with the energy
conditions for a further study. These two datasets are sensitive to
the cosmic evolution in the dark energy-dominated era and turn out
to form good complementary constraints.

\subsection{Hubble parameter}
Observational Hubble parameter data (OHD) generally can be measured
through the differential ages of galaxies
\citep{jimenez2002constraining,simon2005constraints,stern2010cosmic,moresco2012improved,zhang2012four}
and the baryon acoustic oscillation (BAO) peaks in the galaxy power
spectrum \citep{gaztanaga2009}. They have been used in many
cosmological models within the FRW framework
\citep{samushia2006cosmological,stern2010cosmic,zhang2010constraints,zhai2011constraints},
and even the Lema\^{\i}tre-Tolman-Bondi (LTB) void models
\citep{wang2012constraints}. In this paper, we adopt the updated 28
OHD listed in the literature \citep{farooq2013hubble,zhang2012four}.
It is worth mentioning that the latest measurement does not only
extend to the deeper redshift realm, i.e., $H(z=2.3)=224 \pm 8$ km
s$^{-1}$Mpc$^{-1}$ \citep{2013A&A...552A..96B}, but also reveal a
completely new method to obtain the OHD, namely, the BAO peak method
using the Ly$\alpha$ forest.

For the OHD, parameters can be determined by following $\chi^{2}$
statistics
\begin{equation}  \label{OHD statis}
  \chi^{2}_{\textrm{OHD}}(H_{0}, z, \textbf{p}) = \sum_{i} \frac{[H_0 E(z_i) - H^{obs}
  (z_i)]^2}{\sigma_{i}^{2}},
\end{equation}
where $\textbf{p}$ stands for the parameter vector of each dark
energy models. In calculation, we adopt the latest measurement of
$H_0$ by Planck  \citep{collaboration2013planck} as the prior,
$H_{0} = 67.3 \pm 1.2$ km s$^{-1}$ Mpc$^{-1}$.

\subsection{Luminosity distance}

The SNIa sample is widely used for its famous richness. In this
paper, we use the latest supernova Union2.1 compilation of 580
dataset \citep{suzuki2012hubble}. The data are presented as
tabulated distance moduli with errors.  By definition, the
theoretical distance modulus $\mu^{th}$ is the difference between
the apparent magnitude $m$ and the absolute magnitude $M$
\begin{equation}
    \mu^{th}(z)= m(z) - M = 5
    \log D_L(z) + \mu_0,
\end{equation}
where $\mu_0=42.38 - 5 \log h$, and $h$ is the Hubble
constant $H_0$ in units of $100\ \mathrm{km\ s}^{-1}\ \mathrm{Mpc}^{-1}$. The
luminosity
distance function $D_L(z)$ in a flat space can be expressed as
\begin{equation}
    D_L(z) = (1 + z) \int_{0}^{z} \frac{dz'}{E(z'; \textbf{p})},
\end{equation}
where the dimensionless Hubble expansion rate $E(z'; \textbf{p})$ is
the connection between observation and model through Eq.\
(\ref{Eq:expansionrate}). Thus, the $\chi^2$ statistics can be
constructed imitating the Eq.\ (\ref{OHD statis}) but replacing
Hubble parameter as distance modulus. However, in order to abandon
the nuisance parameter $\mu_0$, an alternative way generally can be
performed by marginalizing over it
\citep{pietro2003future,nesseris2005comparison,perivolaropoulos2005constraints}.
An new form of the $\chi^2$ statistics independent of $\mu_0$ is
eventually reconstructed as
\begin{equation} \label{chi2_SN2}
    \tilde{\chi}^{2}(z, \textbf{p})= A - \frac{B^2}{C},
\end{equation}
where
\begin{eqnarray}
     A(\textbf{p}) &=&  \sum_{i} \frac{[\mu^{obs}(z) - \mu^{th}(z; \mu_0 = 0,
     \textbf{p})]^2}{\sigma_{i}^{2}(z)},   \nonumber\\
     B(\textbf{p}) &=&  \sum_{i} \frac{\mu^{obs}(z) - \mu^{th}(z; \mu_0=0,
     \textbf{p})}{\sigma_{i}^{2}(z)},   \nonumber\\
     C &=&  \sum_{i} \frac{1}{\sigma_{i}^{2}(z)}.
\end{eqnarray}
This program has been widely used in the parameter constraint
\citep{wei2010observational},  reconstruction of the energy
condition history \citep{wu2012reconstructing} etc.

\subsection{Energy conditions}
\label{energy condition}

By virtue of the Eqs.\  (\ref{zeropointenergy}), (\ref{total
Friedmann}) and (\ref{total press}), the energy conditions can be
expressed in the below forms. The NEC suggests
\begin{equation} \label{holographic ec}
    1 + \omega \Omega_{\Lambda} \geqslant 0,
\end{equation}
the WEC:
\begin{equation}
    3 M^{2}_{p} H^{2} \geqslant 0 \quad \mbox{and} \quad
   1 + \omega \Omega_{\Lambda} \geqslant 0,
\end{equation}
the SEC:
\begin{equation}
    1 + 3 \omega \Omega_{\Lambda} \geqslant 0 \quad  \mbox{and} \quad
    1 + \omega \Omega_{\Lambda} \geqslant 0,
\end{equation}
the DEC:
\begin{equation}
    3 M^{2}_{p} H^{2} \geqslant 0,  \;
    1 + \omega \Omega_{\Lambda} \geqslant 0, \;  \mbox{and} \;  1 - \omega
    \Omega_{\Lambda} \geqslant 0.
\end{equation}
Because of the non-negativity of $3M^{2}_{p}H^{2}$,  WEC eventually
reduces to NEC.

Assuming the energy conditions, some bounds can be placed on
parameters $\omega$ and $\Omega_{\Lambda}$. Furthermore, the EoS
parameter $\omega$ can be constructed from parameters
$\Omega_{\Lambda}$ (or $\Omega_m$) and $c$, which will be shown in
next section.  Thus, the energy conditions actually provide a series
of constraints on parameters $\Omega_{m}$ and $c$.  Note that the
dark energy component $\Omega_{\Lambda}$ varies with redshift $z$,
and its evolution has been shown in the above section.  This means
that we cannot expect the energy conditions to be constantly
fulfilled or violated.

\subsection{Joint data constraints and energy condition bounds}

\begin{figure}
    \begin{center}
\includegraphics[width=0.45\textwidth]{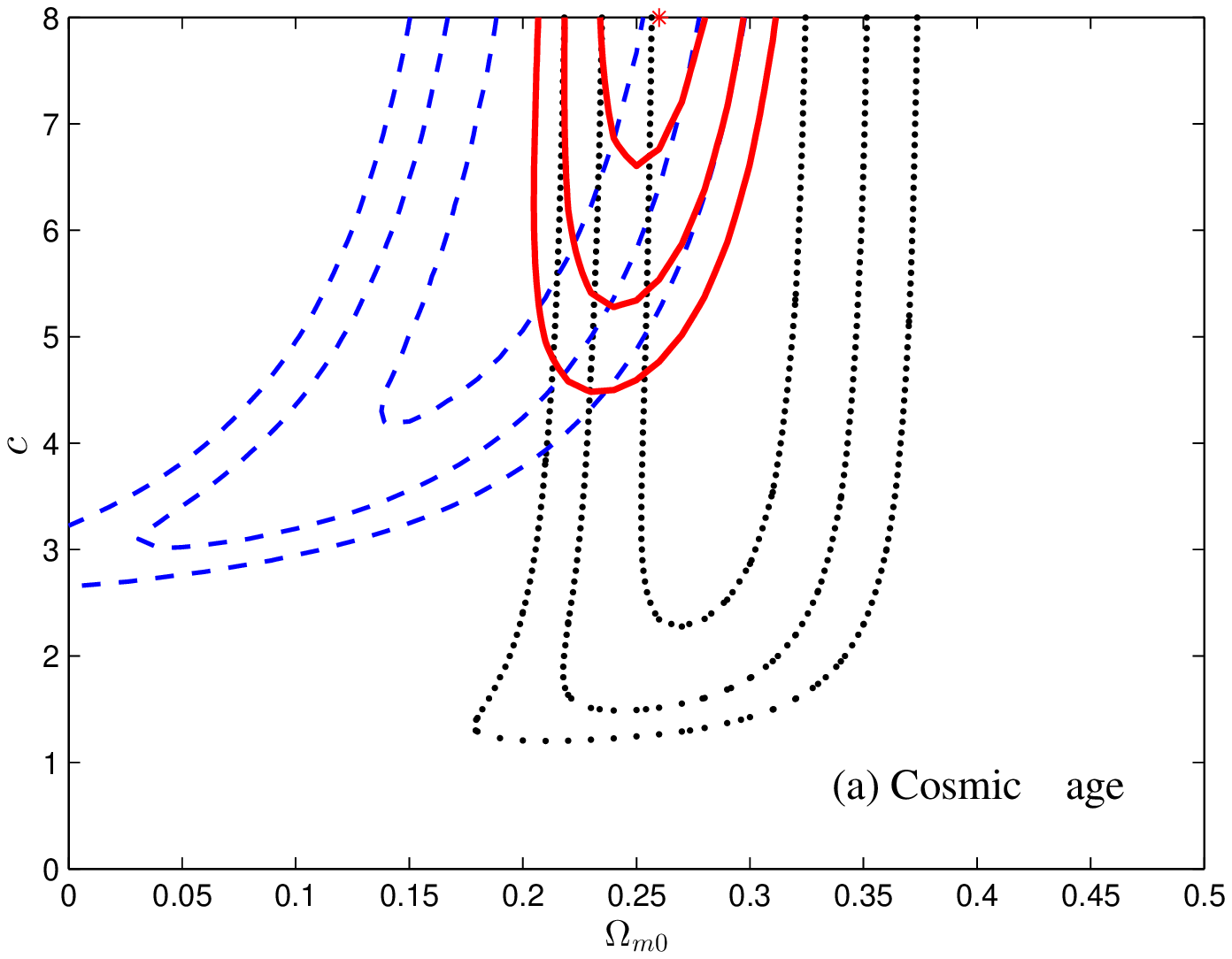}
\includegraphics[width=0.45\textwidth]{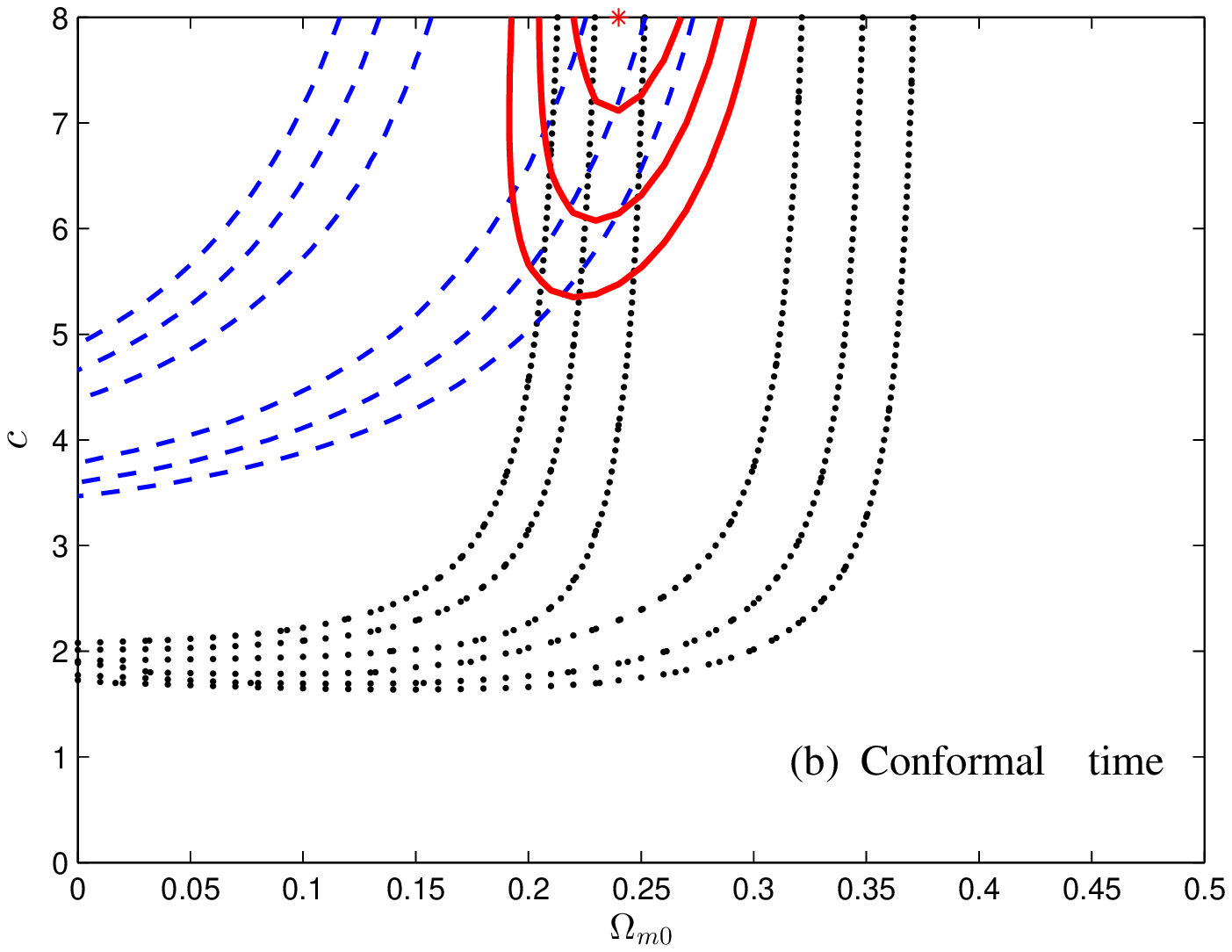}
\includegraphics[width=0.45\textwidth]{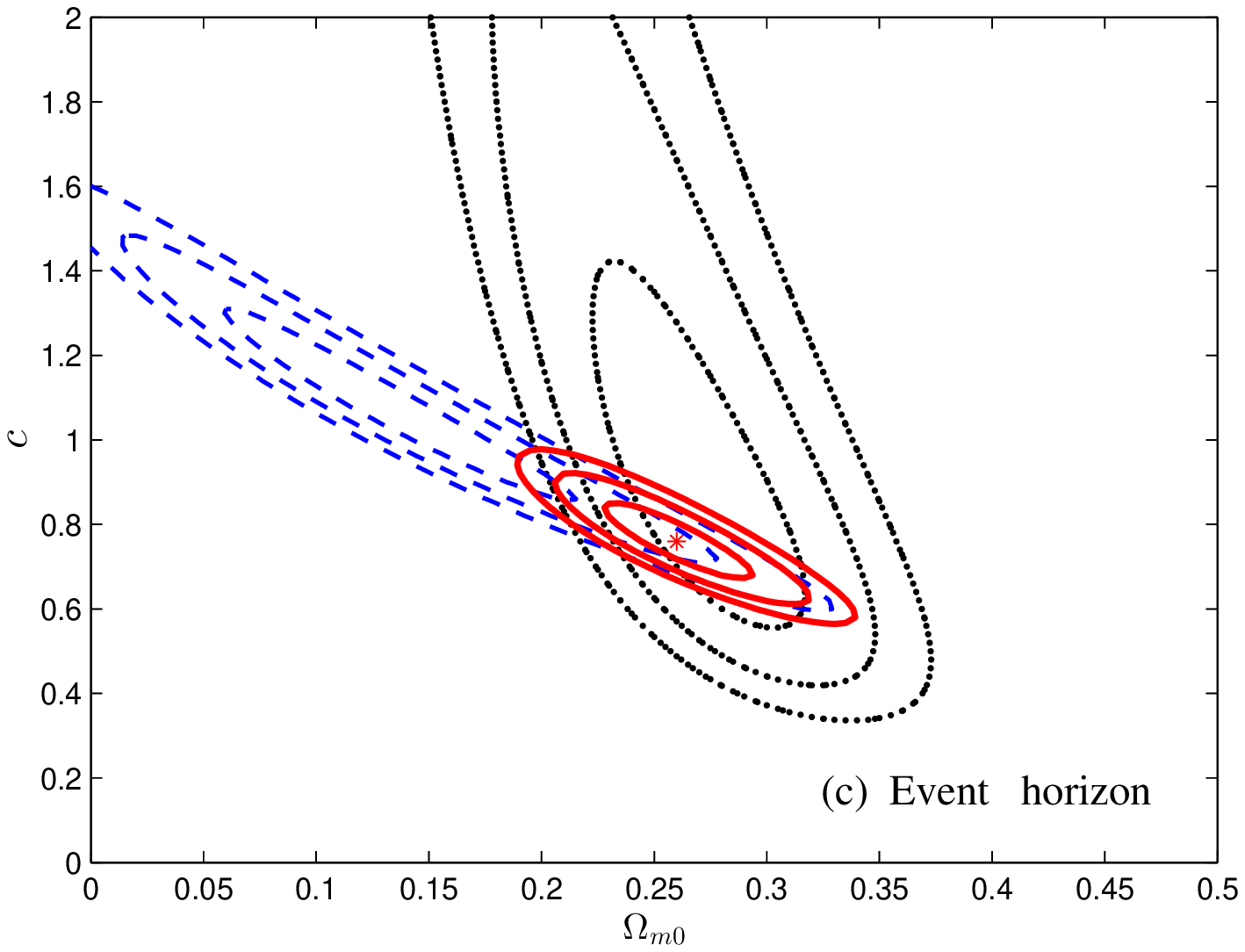}
    \end{center}
    \caption{\label{contour} Constraints on parameters $(\Omega_{m0}, c)$ from
    OHD and SNIa for the three models.  The dashed and dotted contours are
    obtained from individual SNIa and OHD, respectively.  The solid contours are joint
    constraints for OHD and SNIa. The contour levels correspond to 68.3\%,
    95.4\% and 99.7\% confidence regions, respectively. In the calculation of OHD, the latest measurement  $H_{0}=67.3 \pm 1.2$ km s$^{-1}$
Mpc$^{-1}$ \citep{collaboration2013planck} is used as the prior.
The cut-off scenarios are indicated in the figure
    labels (a) to (c).}
\end{figure}

Having carried out the $\chi^2$ analysis with our three models, we
are able to report the resulting constraint on the parameters, as
shown by Figure \ref{contour}.  As can be seen from that figure, the
OHD and SNIa sets display fairly good complementarity, despite some
trouble with bounding the $c$ parameter.  Together they are able to
reduce the parameter degeneracy and give us satisfactory bounds on
$\Omega_{m0}$ for all the three models.  Even if they cannot always
bound the parameter $c$, when used together the data improves our
lower estimate of its possible range by a significant margin.

\begin{figure}
    \includegraphics[width=0.45\textwidth]{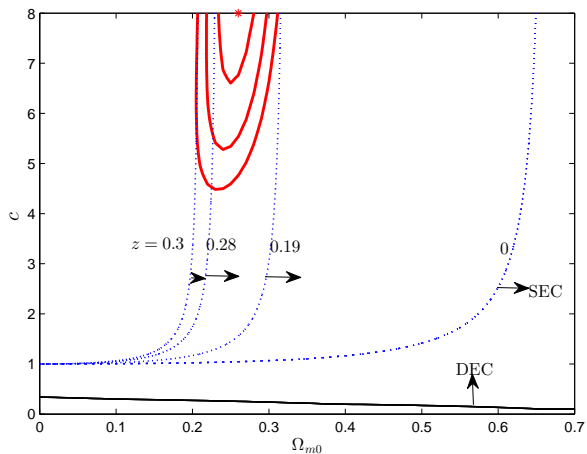}
    \caption{\label{ec_age}  Observational constraints and the energy
    condition bounds for the cosmic age cut-off.  The solid contours correspond
    to the joint constraint from OHD and SNIa data.  Energy conditions bounds are
    labeled by redshift, and arrows point to the direction where the conditions
    are (or use to be) fulfilled.  NEC bounds cover the whole parameter space,
    which is too trivial to be plotted.  Constraint from SEC changes evidently
    with redshift, but constraint from DEC barely changes.  Since constraints
    from the energy conditions are redshift dependent, i.\ e.\  constraints on
    $\Omega_m(z)$, we have transferred the unit into the current parameter
    space $(\Omega_{m0}, c)$ according to the evolution of matter.}
\end{figure}

For the {\em cosmic age cut-off}  model, we find $\Omega_{m0} = 0.26
\pm 0.02$ with $\chi^2_{\textmd{min}}= 592.25$. For a fair
comparison between models, we also consider a conventional criterion
in the literature, i.e., the minimum value in per degree of freedom
$\chi^2_{\textmd{min}}/dof $=0.977. We know that the smaller
$\chi^2_{\textmd{min}}/dof $ is a better choice for models. Assuming
a prior on constant $0<c<8$, we find that the upper limit for $c$ is
absent but $c>6.60$ at 68.3\% confidence level. In fact, it is
consistent with that of the interaction models \citep{chen2011yun}.
We, therefore, could conclude that the spatial structure in form of
cosmic age cut-off is the most essential---regardless of whether
there exists interaction or not between interior matter. On the
other hand, HDE finally can reduce to the standard cosmological
model just because of the fundamental requirement $c \rightarrow
\infty$ in this model. Using the fitted parameters, we estimate the
EoS today of HDE at $\omega_{\Lambda}=-0.994$, very approach to the
cosmological constant. For comparison, bounds required by the energy
conditions at redshift $0 \leqslant z \leqslant 0.35$ are shown in
Figure \ref{ec_age}.  As for the energy conditions, we find that
both NEC and DEC are trivially satisfied by the constrained regions
for the entire redshift range. However, the data are in tension
against SEC fulfillment at redshift $z = 0$.  In addition, we plot
the evolution of the SEC bound at different redshift.  We find that
SEC is fulfilled at redshift $z \gtrsim 0.28$ with $1\sigma$
confidence level, and $z \gtrsim 0.3$ with $3\sigma$ confidence
level. We also note that the observations indicate a violation of
SEC at $z<0.19$.

\begin{figure}
    \includegraphics[width=0.45\textwidth]{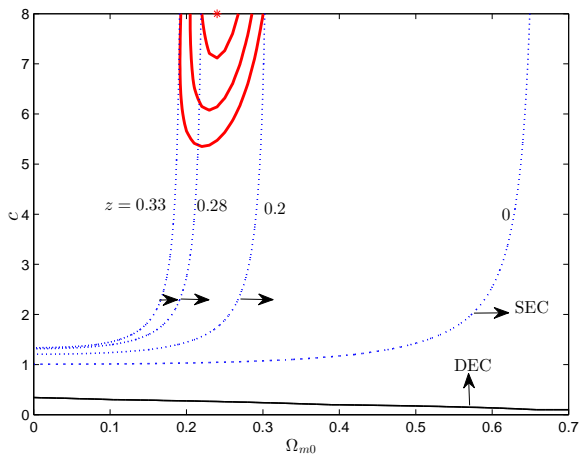}
    \caption{\label{ec_conformal}  Same as Figure \ref{ec_age}
    but for the conformal time cut-off. Analogous results for SEC are
    obtained.}
\end{figure}

For the {\em conformal time cut-off}  model, constraints from data
and the energy conditions are shown in Figure \ref{ec_conformal}.
The joint constraint for OHD and SNIa gives a little lower value of
$\Omega_{m0} = 0.24 ^{+0.03}_{-0.02}$,  a higher constant $c > 7.11$
at 68.3\% confidence level and a worse $\chi^2_{\textmd{min}}/dof=
0.983$. EoS today of HDE in this model is also estimated at
$\omega_{\Lambda}=-0.994$, very approach to the cosmological
constant. NEC and DEC are almost kept stably consistently for the
redshift range. This is same as the situation in age of universe
cut-off. Moreover, an analogous result for SEC is obtained.  That
is, SEC access the valid region at $z\approx 0.2$. However, complete
validity within $1\sigma$ may occur at $z\approx 0.3$.

\begin{figure}
    \includegraphics[width=0.45\textwidth]{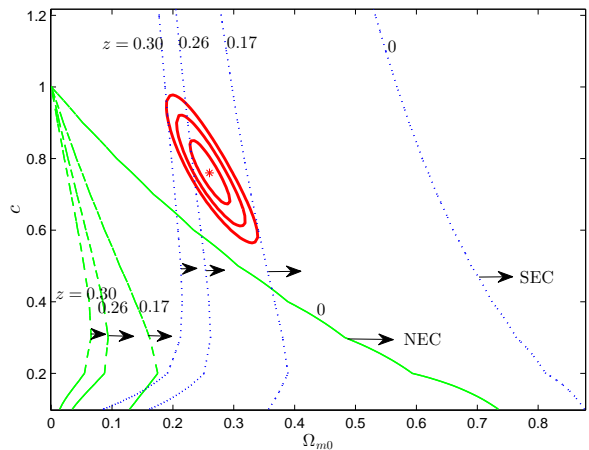}
    \caption{\label{ec_event}  Same as Figure \ref{ec_age} but
    for the event horizon cut-off. Dashed curves show the boundaries of
    constraint from NEC. }
\end{figure}

Finally we perform the $\chi^2$ analysis on the {\em event horizon
cut-off}  model. They present a moderate estimation for matter
density $\Omega_{m0}=0.26 \pm 0.04$ and a closed constraint on $c =
0.76 \pm 0.06$. A similar evaluation as the cosmic age cut-off model
is also available, i.e., $\chi^2_{\textmd{min}} = 592.77$ and
$\chi^2_{\textmd{min}}/dof= 0.978$. Moreover, current equation of
state of dark energy is estimated to be $\omega_{\Lambda}=-1.08 \pm
0.06$, a little deviation from the $\Lambda$CDM model. Comparison
with energy condition bounds is shown in Figure \ref{ec_event}. We
note that NEC can give a much more meaningful bound for this
scenario, especially at $z = 0$. Moreover, we also note that it can
directly and effectively compare with the constraint of OHD. For
SEC, a familiar result is obtained, i.\ e.\ fulfill data at redshift
$z \gtrsim 0.26$ with $1\sigma$ confidence level, and $z \gtrsim
0.30$ with $3\sigma$ confidence level. Most noteworthy is that the
SEC from validity to complete validity within $3\sigma$ only across
the redshift span $0.17 \lesssim z \lesssim 0.30$, which can be used
a good tool to test some models.

\section{Conclusion and discussions}
\label{Conclusions}

We perform $\chi^{2}$ statistics to constrain the holographic and
agegraphic dark energy models using latest OHD and SNIa Union2.1
compilation. The best-fit value of matter density for the three
scenarios almost steadily located at $\Omega_{m0}=0.26$ with
$1\sigma$ confidence level. The main results are shown in Figure
\ref{contour}.  It turns out that $c$ cannot be constrained with a
closed region for the age and conformal time cut-off, which is
consistent with that of the interacting models \citep{chen2011yun}.
We, therefore, could conclude that the spacetime structure in form
of cosmic age cut-off is the most essential---regardless of whether
there exists interaction or not between matter and dark energy.  As
discussed above, these two models eventually approaches the
$\Lambda$CDM model. The EoS today is estimated towards
$\omega_{\Lambda}=-0.994$ for the two models. The constant $c$ is
closed constrained at $c = 0.76 \pm 0.06$ from joint data for the
event horizon cut-off. The EoS in the event horizon cut-off is thus
reconstructed as $\omega_{\Lambda}=-1.08 \pm 0.06$, a little
deviation from the $\Lambda$CDM model. Furthermore, we consider the
evaluation criterion $\chi^2_{\textrm{min}}/dof$ between models. We
find that neither of this several models seem to work very well. By
contrast, the model with conformal time cut-off is the worst.

The energy conditions turn out an useful approach to the study of
cosmology, because of its model independence and universality.  In
our analysis, we superimpose the energy conditions bounds on the
constraints from observational data as done in Figures \ref{ec_age},
\ref{ec_conformal}, and \ref{ec_event} at different redshift. In
these scenarios, we find that SEC is fulfilled at redshift $0.2
\lesssim z \lesssim 0.3$ with $1\sigma$ confidence level. The
observations also require a validity of SEC with $3\sigma$
confidence level at $z>0.3$. Moreover, the SEC is violated by the
observations at $z\lesssim 0.2$. Theoretically, such a small error
could provide a good constraint to future model examination.
Moreover, constraint of the energy condition is physical and
model-independent. Unfortunately, SEC is violated currently, which
is expected in \citep{lima2008energy,wu2012reconstructing}.
Moreover, we find that NEC can provide an additional effective
constraint on $\Omega_{m0}$ and constant $c$ for the event horizon
cut-off,  when it is compared with OHD only. In fact, as far back as
1997, \citet{visser1997energy} has inspected the observation in the
epoch of galaxy formation using the energy condition, and predict an
abnormal matter.

Compared with the work of \citet{wu2012reconstructing}, our results
are obtained from a much narrower class of models.  However, it
should be noted that their work assumed a family of discontinuous
functions as the bases on which the kinematic terms are expanded.
Our result, in contrast, is not subjected to this limitation, but
evolution of the energy condition. Therefore our analysis cannot be
viewed as a special case of theirs. Importantly, our result exactly
determine the fulfilment region of SEC, namely $0.2 \lesssim z
\lesssim 0.3$ for $1\sigma$ confidence level, and is less ambiguous
at higher redshift.

It is possible, according to our results, to distinguish NEC as the
energy condition {\it primus inter pares}. We find that, models that
cannot be well-constrained by the data will also fail by NEC,
yielding too generous a bound.  For the model capable of being
constrained by data, NEC gives a meaningful constraint that is
neither in conflict with, nor entirely implied by the data. In
future analysis of other dark energy proposals, therefore, NEC may
be a worthwhile theoretical consideration before the application of
data analysis.

However, we have to admit a fact that only OHD and SNIa compilations
are used in present paper, although they are the latest ones. If
more sufficient data were considered, such as WMAP,  BAO etc., the
constraint would be tighter and stronger. And power of the energy
conditions would also be disabled. Repeated again, we have
considered the HDE models without interaction according to a fact
\citep{wu2008current}. However, we should also note an important
phenomenon. For the interacting models, degeneracy between matter
density and constant $c$ shown in Figures 2, 3 and 4 of
\citet{chen2011yun} is different from the one of this paper. From
the opposing direction of contour constraints, we have the
possibility to expect that the energy condition could play an
important role in the interacting models. What we can not ignore is
that neither the $\chi^2_{\textrm{min}}/dof$ nor the AIC and BIC
criteria \citep{zhai2011constraints} under current observations rule
out some of these models unambiguously. It is, therefore, reasonable
to expect a powerful future observation to distinguish them.

\acknowledgments We thank the anonymous referee whose suggestions
greatly helped us improve this paper. This work was supported by the
National Science Foundation of China (grant No.\ 11173006), the
Ministry of Science and Technology National Basic Science program
(project 973) under grant No.\ 2012CB821804, and the Fundamental
Research Funds for the Central Universities.

\bibliography{holo}
\end{document}